# Maximum Tolerated Dose Versus Metronomic Scheduling in the Treatment of Metastatic Cancers

Sébastien Benzekry, Philip Hahnfeldt

*Center of Cancer Systems Biology, Steward Research & Specialty Projects Corp., St Elizabeth's Medical Center, Tufts University School of Medicine, Boston, 02135, USA*

## Abstract
Although optimal control theory has been used for the theoretical study of anti-cancerous drugs scheduling optimization, with the aim of reducing the primary tumor volume, the effect on metastases is often ignored. Here, we use a previously published model for metastatic development to define an optimal control problem at the scale of the entire organism of the patient. *In silico* study of the impact of different scheduling strategies for anti-angiogenic and cytotoxic agents (either in monotherapy or in combination) is performed to compare a low-dose, continuous, metronomic administration scheme with a more classical maximum tolerated dose schedule. Simulation results reveal differences between primary tumor reduction and control of metastases but overall suggest use of the metronomic protocol.

## Introduction
In cancer treatment, scheduling of anticancer agents can impact on the overall outcome of the therapy, even with the same total amount of administered drug. Indeed, changing the temporal administration protocol influences the dynamics of the system, in a nonlinear fashion. In the last decade, clinical and preclinical efforts were engaged to develop novel therapeutic approaches for scheduling of cytotoxic (CT) agents as proved by the growing field of metronomic chemotherapy (1,2,3,37) - the chronic administration of chemotherapy at relatively low, minimally toxic doses on a frequent schedule of administration, at close regular intervals, with no prolonged drug-free breaks. This approach opposes to the classical way of administering CT, designed here by maximum tolerated dose schedules (MTD), which gives the largest possible amount of drug at the beginning of the cycle and then lets the patient recover from toxicities. Experimental and clinical studies have been performed in order to determine the best scheduling of CT agents (3–5), but there is still no clear answer about optimal temporal administration protocols.

Besides of well-established cytotoxic molecules, anti-angiogenic (AA) agents are currently under investigation. These therapies are designed to target the tumor vasculature rather than the cancerous cells themselves, based on the discovery in the 70's of the angiogenesis importance in a tumor's development (6). It was then evidenced that tumor growth is strongly dependent on development of vascular support, which a tumor is able to impulse by emission of stimulating molecules such as Vascular Endothelial Growth Factor (VEGF) and balance with endogenous inhibitory agents such as endostatin, angiostatin or thrombospondin-1 (7,8). Targeting angiogenesis appeared then as a promising idea but up to now only few molecules could obtain approval (bevacizumab, a monoclonal antibody targeting VEGF and sunitinib, a tyrosine kinase inhibitor binding on VEGF endothelial cells' receptors are two examples). Appearance of these new drugs rises the question of their optimal scheduling and oncological literature contains studies addressing this issue (9–11), but the question is still open.

Mathematical modeling offers a relevant theoretical framework in which studying these concerns. Cancer modeling has a long history in the 20th century with most of the efforts being targeted towards modeling of tumor growth (see (39) for a review). However, metastasis is the main cause of death in a cancer disease (12). In 2000, Iwata et al. (13) introduced a mathematical model for the development of the population of metastases, which was then further studied in (14,15) in particular in the intent to incorporate effect of the chemotherapy. On the other hand, Hahnfeldt et al. (8) developed a phenomenological model for interactions between a tumor and its vasculature, allowing to account for anti-angiogenic therapy. We coupled these two approaches in (16,17) and obtained a global model of temporal progression of a cancer disease, written at the organism scale and taking into account for the main processes of the pathology: proliferation, angiogenesis and metastatic spreading. The model incorporates effects of systemic AA and CT treatments and illustrations of its clinical relevance were given in (18).

Based on mathematical models of cancer growth, optimal control theory has been applied in numerous studies (see (38) for a review), starting with administration of chemotherapy acting only on cancer cells (see for instance the work of Swan (19,20)). Problems in this context arise from tumor heterogeneity as the cancerous cells population comprises subpopulations with different drug sensitivities, either due to their position in the cell cycle or to different degrees of acquired resistance to the cytotoxic drug. A possible mathematical approach is to use discrete compartments for different subpopulations, for instance quiescent and proliferative or subpopulations having different sensitivities to the drug (40), or both (48). Others use a continuous variable to describe progression within the cell cycle (42,43), which allows to model action on transition rates between phases. An optimal control problems integrating pharmacokinetics (PK) and pharmacodynamics (PD) considerations for chemotherapy is analyzed in (21). Toxicity on healthy cells is a major concern and often appears as a constraint in the optimization problem. For example, the Model 1 project (23–27) drove a clinical phase I study by a mathematical model focused on hematotoxicity of the chemotherapies. The optimization schedule computed by the model allowed densification of a standard protocol while dynamically controlling the toxicities. Models developed in (41,42,43) also deal with optimization problems focused on reduction of toxicity for healthy tissues. Genuine use of different circadian synchronization between healthy and cancer tissues is theoretically studied in (53) and an experimentally-validated PK/PD model for optimization of a cytotoxic drug used for treatment of colorectal cancer is designed in (54,55) to practically optimize circadian delivery of the drug. For AA therapy, using the tumor growth model of Hahnfeldt et al. (8) and its further refinements and analysis proposed in (47) optimal control problems have also been widely investigated. Optimal schedules for AA treatments alone perturbating tumor growth have been studied in (48) and more extensively by Ledzewicz and Schättler in (28–31). Combination of radiotherapy and an AA drug is studied in (32), using a simplification of the Hahnfeldt model. Combination of CT and AA therapy has been considered in (33). However, as expressed before, these models do not take into account the metastatic development of a cancer disease.

In this paper, we formalize an optimal control problem for the metastases and present numerical simulations of the effect of the scheduling strategy on the cancer disease. They demonstrate the importance of scheduling for anticancer agents and by comparing minimization objectives defined on the primary tumor and on the metastases, we study the differences between primary tumor reduction and control of the metastatic spreading. A brief mathematical analysis of the theoretical optimal control problem is presented in the appendix.

## Formulation of an optimal control problem for the metastases
*Primary tumor*
For primary and secondary tumor dynamics, we will assume that the growth law is given by the Hahnfeldt (8) model, modified by the action of a therapy. We denote by $V_p$ the volume (expressed in mm³ for instance) of the primary tumor and by $K_p$ its carrying capacity (same unit as the volume, assumed to represent the vasculature state of the tumor), grouped in a global variable $X_p(t) = (V_p(t), K_p(t))$ for the primary tumor state. The treatment is denoted by $u(t) = \begin{pmatrix} C(t) \\ A(t) \end{pmatrix}$ with $C(t)$ and $A(t)$ the dose rates of CT and AA drugs respectively. Dynamics of the system are given by

$$\dot{X}_p = \bar{G}(X_p; u), \quad \bar{G}(X; u) = G(X) - B(X)u(t)$$

$$G(X) = G(V, K) = \begin{pmatrix} aV \ln\left(\dfrac{K}{V}\right) \\ cV - dV^{2/3}K - \lambda K \end{pmatrix},$$

with $B(X) \in \mathcal{L}(\mathbb{R}^2, \mathbb{R}^2)$ a matrix describing how does the treatment act on the tumor. For example $B(X) = \begin{pmatrix} 0 & 0 \\ 0 & eK \end{pmatrix}$ for an AA drug alone. We assume in a first approximation that the input flows of the drugs are the same than the efficient concentrations acting on the tumor, thus neglecting the role of pharmacokinetics and pharmacodynamics. This could be dealt with by replacing $u(t)$ by $E(t, u(t))$ with $E$ being a (possibly nonlinear) function describing the effects on cancer of the dose rates $(C(t), A(t))$ on the tumor and vascular compartments.

We will consider two objectives to be minimized for tumor growth : the tumor size at a fixed end time $T$ and the maximal tumor reduction during the time interval $[0,T]$. We denote
$$J_T(u) = V_p(T;u) \quad \text{and} \quad J_m(u) = \min_{t \in [0,T]} V_p(t;u).$$
A minimization problem on the primary tumor (studied in (33) for a free end time) then writes: find $u^* \in \mathcal{U}_{ad}$ such that $J_m(u^*) = \min_{u \in \mathcal{U}_{ad}} J_m(u)$ with $\mathcal{U}_{ad}$ being the space of admissible controls, integrating toxicity constraints (see below for its expression). A similar problem is obtained by changing $J_m$ into $J_T$.

*Metastases*
Although the problem of best reduction of the primary tumor size is of great relevance in clinical practice, the metastatic state cannot be neglected due to its importance in a cancer disease and its implications in the possibility of relapse. Two practical examples of clinical situations where optimization of scheduling is relevant and metastases have to be taken into account could be the followings. In metastatic breast cancer for instance, after primary tumor resection, the clinician wants to control the number of metastases above a given size, for large time. He wants to give a combined CT - AA treatment such that in the next years no visible metastasis appears. In the context of metronomic CT that has the advantage to induce weaker hematological toxicities and thus does not require intricate modeling for this matter (as done in (23) for instance). The time horizon is then of the order of one year and resistances developed by cancerous cells have to be taken into account. Number of metastases and their sizes have to be kept under control.

In a first attempt to theoretically study the involved dynamics and for computational commodity as well as comparison with previous other theoretical studies on the primary tumor, in particular (29), we will place ourselves in a framework where the time span is thought as being a therapy cycle, thus of the order of weeks.

We use the model developed in (16,17) for the metastatic development. It is based on the combination of the model first introduced by (13) and the Hahnfeldt model (8). Both of these mathematical descriptions were proven to be able to describe biological data (clinical data of a patient with hepatocellular carcinoma for the first one, Lewis lung carcinoma in mice for the second). The overall philosophy is to think at the organism scale and to write a systemic model for the development of the metastatic colonies. While (13) only considered size structure and neglected angiogenesis, we take this process into account by using as growth law for each tumor the two-dimensional model of (8). The main variable of the model is the density of metastases $\rho(t, X; u)$ structured by the trait $X = (V, K)$ with $V$ being the tumor volume and $K$ the carrying capacity. A balance law leads to the following structured partial differential equation (see (16) for more detailed modeling concerns):

$$\begin{cases} \partial_t \rho(t,X;u) + \text{div}(\rho(t,X;u)\overline{G}(X;u)) = 0 & ]0,T[\times\Omega \\ -\overline{G}(t,\sigma;u)\cdot v(\sigma)\rho(t,\sigma;u) = N(\sigma)\left\{\int_\Omega \beta(X)\rho(t,X;u)dX + \beta(X_p(t;u))\right\} & ]0,T[\times\partial\Omega \\ \rho(0,X;u) = \rho^0(X) & \Omega \end{cases}$$
(1)

where

$$\Omega = \left]V_0, \left(\frac{c}{d}\right)^{3/2}\right[^2, \quad N(\sigma) = \delta_{\{\sigma=(V_0,K_0)\}}, \quad \beta(V,K) = mV^\alpha$$

with $\delta$ the Dirac measure, $V_0$, $K_0$ the vasculature of a metastasis at birth and $m$ and $\alpha$ two parameters of metastatic emission. Mathematical analysis of the problem (1) without therapy has been performed in (17). Theoretical analysis with therapy as well as numerical analysis of an approximation scheme for the simulation of (1) was performed in (16).

Toxicity is dealt by imposing constraints on $u$. We don't include a precise description of hematological toxicities. Common toxicities (like renal ones for instance) are taken into account by imposing, similarly as in (33) : a) maximal local values for $C(t)$ and $A(t)$ denoted by $c_{max}$ and $a_{max}$ respectively, which are non-negative constants and b) maximal total amounts of drug delivered (corresponding to the clinical Area Under the Curve (AUC)), $C_{max}$ for the CT and $A_{max}$ for the AA, again two non-negative constants. We consider thus the following space of admissible controls:

$$\mathcal{U}_{ad} = \left\{ u \in (L^\infty(0,T))^2; \begin{pmatrix} 0 \\ 0 \end{pmatrix} \leq u(t) \leq \begin{pmatrix} c_{max} \\ a_{max} \end{pmatrix} \forall t \text{ and } \int_0^T u(t)dt \leq \begin{pmatrix} C_{max} \\ A_{max} \end{pmatrix} \right\}.$$

We will consider two objectives to be minimized for the metastases: the total number of metastases and the total metastatic mass, at the end time. Their expressions are given by

$$J(u) = \int_\Omega \rho(T,V,K;u)dVdK \quad \text{and} \quad J_M(u) = \int_\Omega V\rho(T,V,K;u)dVdK$$

The optimal control problems that we will consider on metastatic dynamics are: find $u^* \in \mathcal{U}_{ad}$ such that $J(u^*) = \min_{u \in \mathcal{U}_{ad}} J(u)$ and the equivalent with $J_M$ instead of $J$. A short mathematical analysis of this problem is presented in the appendix containing proof of the existence of an optimal control and derivation of a first order optimality system.

## Numerical simulations of a simplified optimization problem. Scheduling strategy. Metronomic versus MTD schedules

A natural biological question is to know if the solutions to the optimization problems defined above differ between primary tumor and metastases?

The answer to this question is no, as illustrated by numerical simulations in this section. Heuristically it makes sense since one can imagine a scenario having different effects on the growth of each tumor and on the total number of metastases at the end: if we let tumor growth being important during a large time and give a large amount of drug at the end, the tumors can be largely reduced whereas the total number of metastases is still high since during the whole time where growth was important there was more metastases emission and the final decrease of tumors sizes does not impact a lot on this already important spread.

As suggested in (29), an easy-to-handle but nevertheless clinically relevant situation is to look at optimality in a two-dimensional framework where the problem is to administer total fixed amounts of agents $(C_{max}, A_{max})$ from time 0 to times $(t_C, t_A)$ at constant rates $\bar{C} = \frac{C_{max}}{t_C}$ and $\bar{A} = \frac{A_{max}}{t_A}$, and then set the control to zero. This means that the set of admissible controls is

$$\mathcal{U}_{ad} = \left\{ u \in (L^\infty(0,T))^2 ; C(t) = \bar{C}\mathbf{1}_{[0,t_C]}(t), A(t) = \bar{A}\mathbf{1}_{[0,t_A]}(t), \left(\frac{C_{max}}{t_C}, \frac{A_{max}}{t_A}\right) \leq (c_{max}, a_{max}) \right\}$$

Notice that now the constraints on local maximal values become constraints on minimal values for $t_C$ and $t_A$ that become the optimization variables. We will refer to them as the scheduling strategies and write $J(t_C, t_A)$, $J_T(t_C, t_A)$ and $J_m(t_C, t_A)$ instead of $J(C,A)$, $J_T(C,A)$ and $J_m(C,A)$. In the monotherapy cases, we won't write the dependency on the other drug. The questions we consider are the following.

Is the best anti-cancer efficacy achieved by the lowest values of $t_C$ and $t_A$ (intense MTD protocol) or rather by the highest values of $t_C$ and $t_A$ (continuous low dose metronomic protocol)? Is there a non-trivial optimum between these two situations? Is there a difference in the optimal strategy for the four considered objectives?

In (29), the situation of AA monotherapy (i.e. $c_{max} = 0$) was investigated on the primary tumor. Our aim here is to extend this approach by looking at the behavior on the metastases and also combination of CT and AA therapy. The values of the parameters that we use for the tumor growth are the same as (29) and mostly come from (8) where they were fitted to growth curves of Lewis lung carcinoma in mice

$a = 0.0084$ day$^{-1}$, $\lambda = 0.02$ day$^{-1}$, $c = 5.85$ day$^{-1}$, $d = 0.00873$ day$^{-1}$mm$^{-2}$

For the effect of the treatments we take

$$B(X) = \begin{pmatrix} \gamma(V - V_0) & 0 \\ 0 & \phi(K - V_0) \end{pmatrix}, \quad \gamma = 0.15 \ day^{-1}, \quad \phi = 0.1 \ day^{-1}$$

For the metastatic emission parameters we use $m = 0.001$ mm$^{-2}$ and $\alpha = 2/3$. Concerning the initial conditions, we take the ones corresponding to the simulation of the model after 40 days starting with an initial tumor of size $10^{-6} mm^3$ (= 1 cell) and carrying capacity 625 mm³ (value taken from (8)). This gives for the primary tumor $(V_{0,p}, K_{0,p}) = (1015, 6142)$ and some non-zero initial condition $\rho^0$ for metastases. The newly created metastases enter the system with $(V_0, K_0) = (10^{-6}, 625)$. We run the simulations during a total time $T = 10$ days and take $A_{max} = 300$ (consistently with the order of the total doses administrated in (8)) and $C_{max} = 30$. For the maximal local values we arbitrarily take $c_{max} = 7.5$ and $a_{max} = 75$ so that the minimal administration duration for both drugs will be 4 days. Although this number is not clinically relevant as chemotherapy in MTD schedules could be administered all in one day (see for instance (34)), the actual value is not of strong importance since we are interested in studying the effect of condensed doses of agents versus protracted doses. All the results that we present here are qualitatively the same if the minimal scheduling strategy is set to 1 day. We will refer to the most condensed strategy ($t_C$ or $t_A$ equal to 4) as the Maximum Tolerated Dose (MTD) strategy and the most protracted ($t_C$ or $t_A$ equal to 10) as the metronomic strategy.

The numerical simulations were performed using an approximation scheme developed in (16). It is a Lagrangian scheme based on a change of variable used to straighten the characteristics.

In all the presented Figures, the scale is only valid for one objective (most of the time $J_m$) and the other curves have been rescaled to fit in the same plot.

*AA monotherapy*

We first investigate the case of AA monotherapy and so take $c_{max} = 0$. Looking at the two extreme situations on the tumor evolution of giving the whole dose during a small time ($t_A = 4$) or rather during a large time ($t_A = 10$) (see Figure 1), we already observe that the two strategies have a complete different result concerning the tumor size at the end of the simulation, with a better effect of the second one. This fact had already been observed in (35).

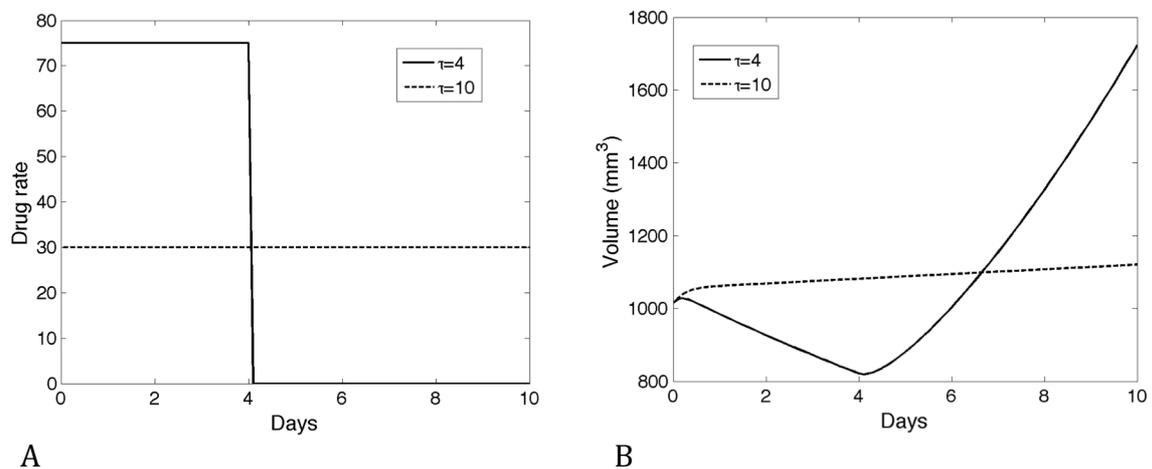

Figure 1: Two extreme examples of delivery of the AA drug on the tumor evolution. A. Drug profile. B. Tumor size

In Figure 2 are plotted simulation results showing the four objectives $J_m, J_T, J$ and $J_M$ over the therapeutic strategy $t_A$.

The first observation is that all the objectives are non-constant. This shows that changing the scheduling strategy of the drug has an impact on its efficacy : different scheduling give different efficacies, for all the objectives.

Let us focus first on the two tumor objectives $J_m$ and $J_T$. They have opposite behaviors. Figure 2.A shows that what minimizes $J_m$ (black squares curve) is the MTD strategy which gives the therapy in a strong dose/short time way. Then, as $t_A$ becomes larger tumor reduction is worse, reaching eventually a plateau due to the fact that the tumor was not reduced on $[0,T]$, the minimal value on this interval being then the initial one. In clinical practice, the physician is often looking for the most immediate results on tumor reduction, as being tangible elements of the response of the patient to the treatment. Moreover, having the sharpest decrease of tumor mass is believed to minimize development of resistances. These rationales lead to try to minimize objective $J_m$ and this first finding which shows that $J_m$ is best minimized by the MTD strategy is in accordance with the classical use of chemotherapeutic agents (although the results shown here are for AA therapy, we obtain the same behavior when simulating CT monotherapy, see below). However, this result is not robust when varying the initial condition and for some values of $V_{0,p}$ and $K_{0,p}$, $J_m$ is best minimized by a scheduling strategy $t_A^*$ different from the MTD (see (29) for an example). This is due to an intricate structure of the solution to the function-valued optimal control problem, studied in (30) where it is shown to exhibit singular portions for some values of initial conditions and parameters.

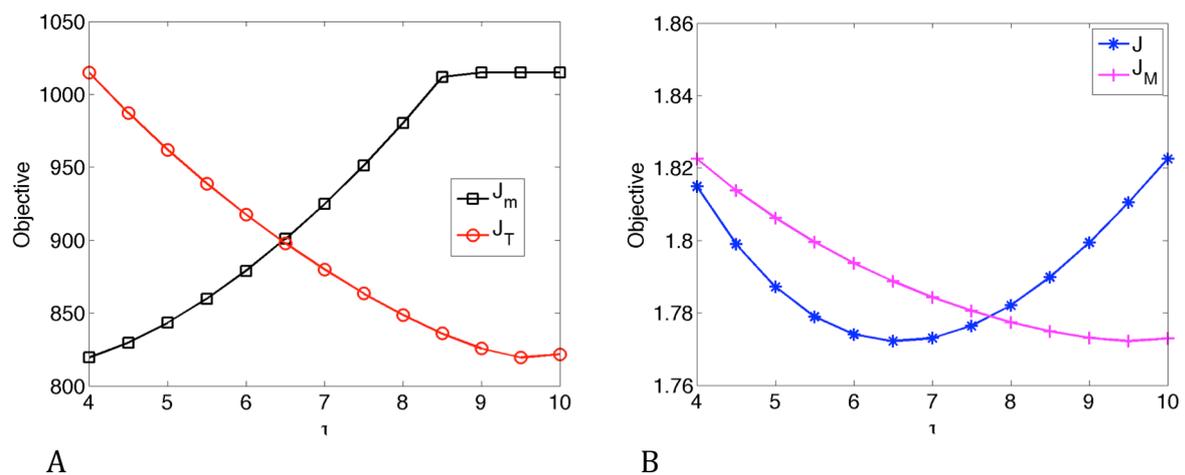

Figure 2: Anti-angiogenic monotherapy. A. Two criteria on the primary tumor : tumor size at the end $J_T$ and minimal tumor size $J_m$ (scale only valid for $J_m$). B. Two criteria on metastases : total number of metastases J, the metastatic mass $J_M$ (scale only valid for J).

Objective $J_T$, the size of the tumor at time $T$ represented by the red circles curve in Figure 2.A, exhibits the complete opposite behavior. It is a decreasing function of $t_A$ (except for the last point, which is an artefact coming from the endpoint of the simulation being the same as $t_A$ in this last simulation). This suggests that the best strategy for this objective would be the metronomic one which gives the therapy in a low dose/large time way. Hence, according to the dynamics of the primary tumor given by the Hahnfeldt model, minimizing $J_m$ or $J_T$ yield to two opposite scheduling strategies, MTD for $J_m$, metronomic for $J_T$.

The metastatic objectives give also two different answers. First, the total number of metastases $J$ (blue stars curve) suggests a different strategy than the tumor objectives. It is a convex function of $t_A$ that has a non-trivial minimal value at $t_A^* = 6.5$ days. This means that the best strategy in order to limit metastatic spread would be neither the MTD nor the metronomic one, but some optimal value of $t_A$ between the two extremes. This result underlines the importance of taking into account for metastases in the clinic, as the best way to cure the primary tumor differs from the best way to reduce the number of metastases. However, the metastatic mass

$J_M$ curve (purple crosses curve) is almost the same as the $J_T$ curve. This could be surprising since we could expect the overall behavior of $J_M$ to result from a mix between the one of $J$ and $J_T$. To understand better what happens here, we have to think about the metastatic mass as the product of the mean metastatic size and the number of metastases, namely

$$(2)\; J_M(t_A) = \int_\Omega V \rho(T,V,K;t_A) dV dK = \underbrace{\frac{\int_\Omega V \rho(T,V,K;t_A) dV dK}{\int_\Omega \rho(T,X;t_A) dX}}_{\text{Mean size } \bar{V}(t_A)} \cdot \underbrace{\int_\Omega \rho(T,X;t_A) dX}_{J(t_A)}$$

Then the variations in $J_M$ due to changes in the scheduling are the product of variations of the mean size $\bar{V}(t_A)$ and $J(t_A)$. From studying the effects on the primary tumor, we obtained (results not shown) that for the size at the end, whatever the initial size is, the optimal strategy is always the metronomic one. Hence the mean size $\bar{V}(t_A)$ is also best reduced by this strategy. Let us now expose an heuristic argument which explains why the mean size wins the competition with the number of metastases in expression (2). On the interval where $J$ is decreasing, both functions $\bar{V}$ and $J$ are decreasing and the minimum is reached at the right end of the interval, thus we can restrict ourselves to some interval $[\tau, 10]$ where $J$ is increasing and $\bar{V}$ is decreasing (in Figure 2.B, $\tau \simeq 7.5$). In first approximation of both functions being linear on this interval, simple calculations show that the product of the two would be a concave function, hence reaching its minimal value at one of the endpoints of the interval, i.e.
$\min_{t_A} J_M(t_A) = \min\left(\bar{V}(\tau)J(\tau), \bar{V}(10)J(10)\right)$. What determines then which endpoint has the smallest value is the relative ratios of increase for $J$ and decrease for $\bar{V}$ on the interval $[\tau, 10]$, and not their absolute value. Here this writes

$$\frac{J_M(\tau)}{J_M(10)} = \frac{\bar{V}(\tau)}{\bar{V}(10)} \frac{J(\tau)}{J(10)} \simeq 1.1650 \times 0.9747 = 1.1355 > 1$$

and this explains why the minimum value is reached for $t_A = 10$, because the relative reduction due to scheduling was more important for the mean size than for the number of metastases. This argument is only valid when $J$ and $\bar{V}$ are linear on $t_A = 10$, which is not exactly the case in the Figure 2, but seems to be valid as first approximation.

In the Table 1 are indicated the amplitude of variation rates of the four objectives. They are defined as follows: for each value of $t_A$, we compute the relative difference between the value of the objective with strategy $t_A$ and the value of the underlying quantity at the beginning of the simulation, before the treatment. For instance, for $J_m$ the formula is

$$(3)\ I(t_A) = \frac{J_m(t_A) - V_{0,p}}{V_{0,p}}$$

The amplitudes indicated are then the minimal and maximal values of these relative differences ($\min_{t_A} I(t_A)$ and $\max_{t_A} I(t_A)$). This is a measure of how effective the treatment has been relatively to this objective and what is the difference between the worst and best therapeutic strategy. For example the variation rates for $J_m$ show that the best tumor reduction was 19% with the MTD strategy and no reduction with the metronomic one. Notice that, in the construction of the model the number of metastases $J$ will always be increasing, as metastases don't go out of the system. However, the therapy impacts on how much it will increase. Notice also that the values of these variation rates depend on the parameters used for the simulations. For instance as seen in the Table 1 for $J_T$, for this value of the total administered dose $A_{max}$ there was never reduction of the tumor at the end time comparing to the start time. With a stronger $A_{max}$ we would obtain negative variation rates, but we used $A_{max} = 300$ since we think about the AA drug as a cytostatic drug, willing to stabilize the growth of the tumor rather than reduce it, at least on small time scales. It is more interesting to look at the amplitudes. For instance for $J_M$ we go from a 33% augmentation to a 154% augmentation between the best and worse strategies, emphasizing the importance of the scheduling of the drug in the treatment.

| Objective | $J_m$ | $J_T$ | $J_T$ | $J_M$ |
|---|---|---|---|---|
| Variation rates | -0.19/0.00 | 0.10/0.70 | 1.32/1.38 | 0.33/1.54 |

Table 1: Variation rates of the objectives

In conclusion, the metastatic mass index as well as $J_T$ suggest that the best strategy for AA drug is to deliver it at low doses during a long time, consistently with preclinical results obtained in (9,10,36) and other simulation study of AA scheduling for the primary tumor in (35). Objective $J_m$ is best minimized by the opposite, MTD strategy and the number of metastases $J$ exhibits a nontrivial minimum. The fact that the best strategy is the same for $J_M$ and for $J_T$ and is different from the best strategy for $J$ means that what best reduces the metastatic mass is a situation with a lot but small metastases, which is preferred to the reverse situation with few but bigger metastases.

*CT monotherapy*

## Scheduling

Results from simulations of cytotoxic monotherapy ($a_{max} = 0$) are represented in Figure 3.

The tumor objectives $J_m$ and $J_T$ give both similar answers as in the AA monotherapy case of Figure 2. However, the total number of metastases $J$ exhibits a different qualitative behavior than previously. While $t_A \mapsto J(t_A)$ was a non-monotonous convex curve for AA monotherapy, here $t_C \mapsto J(t_C)$ is an increasing map. This fact was robust when varying the total amount of given agent $C_{max}$. Hence, the best scheduling strategy for $J$ is different for AA monotherapy (nontrivial optimal value $t_A^*$) than for CT monotherapy (MTD

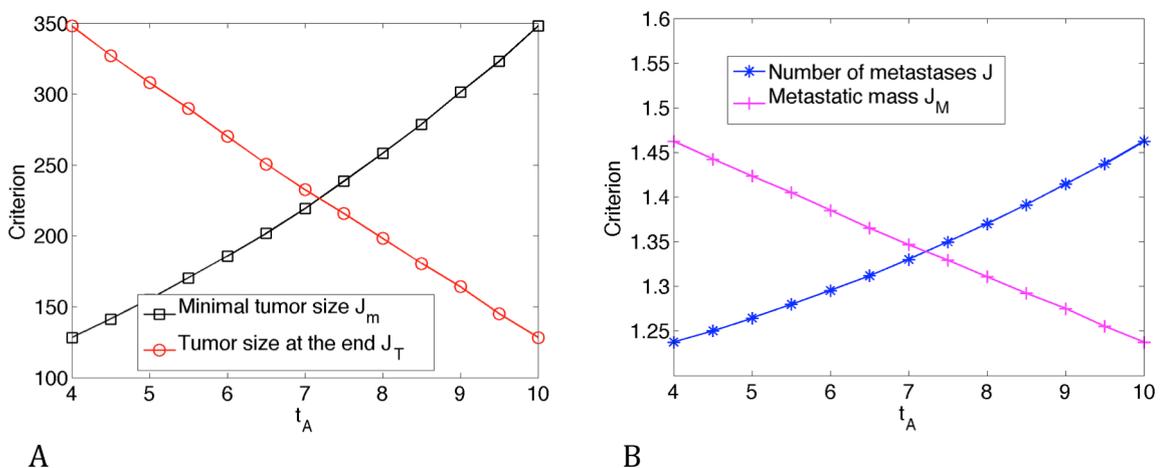

Figure 3: Cytotoxic monotherapy. A. Two criteria on the primary tumor : tumor size at the end $J_T$ and minimal tumor size $J_m$ (scale on valid for $J_m$). B. Two criteria on metastases : total number of metastases J, the metastatic mass $J_M$ (scale only valid for J)

strategy). On the opposite, the metastatic mass objective behaves the same as for AA monotherapy, it is best minimized with the metronomic strategy. The same argument as above explains this fact, i.e. relative variation in the mean size is more important than relative variation in the number of metastases.

**Influence of the parameter m on the metastatic objectives**

With the value of $m$ chosen before and such a small end time, almost all the metastases were emitted by the primary tumor, this amount being given by

$$(4) \int_0^T \beta(V_p(t))dt = m\int_0^T V_p(t)^\alpha dt.$$

Hence, reducing the number of metastases was equivalent to reducing this last expression and we could think that this would always be the case: regarding to scheduling optimization, the best strategy for $J$ could always be the same as the best strategy for the integral expression (4) and we wouldn't need to simulate the whole model given by (1) to study the effect of scheduling on the number of metastases. However, as shown in Figure 4, this is not the case anymore for large values of $m$.

For these simulations, we did not consider any initial condition for the metastases and took $\rho^0 = 0$. In Figure 4.A, we observe that the curves for $J$ and $\int_0^T \beta(V_p(t))dt$ are almost identical and could conclude that the scheduling reducing the best the metastases is the one reducing the best $m\int_0^T V_p(t)^\alpha dt$. But for large values of $m$ this is not the case anymore since the metastases curve has a non extremal minimizer, as illustrated in the Figure 4.B, while of course the integral expression has the same shape since it is linear in $m$.

Changing the value of $m$ affected the shape of the curve for $J$, causing the optimal strategy to move to the right. But it did not affect the metastatic mass, which could be surprising. Indeed, since metastatic mass results from a balance between size and number of metastases, we could expect that increasing the number of metastases by increasing the strength of the spreading (parameter $m$) would influence the behavior of the metastatic mass $J_M$ towards the one of the number of metastases $J$. We could then think that the optimal strategy for $J_M$ would become closer to the optimal strategy for $J$. As shown in Figure 4, the qualitative behavior of $J_M$ and optimal strategy remain the same (metronomic) whatever $m$ being small or large (and also for the intermediate values, not shown here). Indeed, as heuristically explained above, the global outcome for the metastatic mass depends on the balance between the reduction ratios of the mean size $\bar{V}$ and the number of metastases $J$. In our simulations, when $m$ ranged from 0.001 to 100, these ratios ranged respectively from 0.38 to 0.53 and 0.68 to 0.89. Hence, the mean size always wins the competition. In clinical terms, whatever would be the intrinsic metastatic aggressiveness of a patient's pathology, the strategy reducing the best the metastatic mass is the metronomic one.

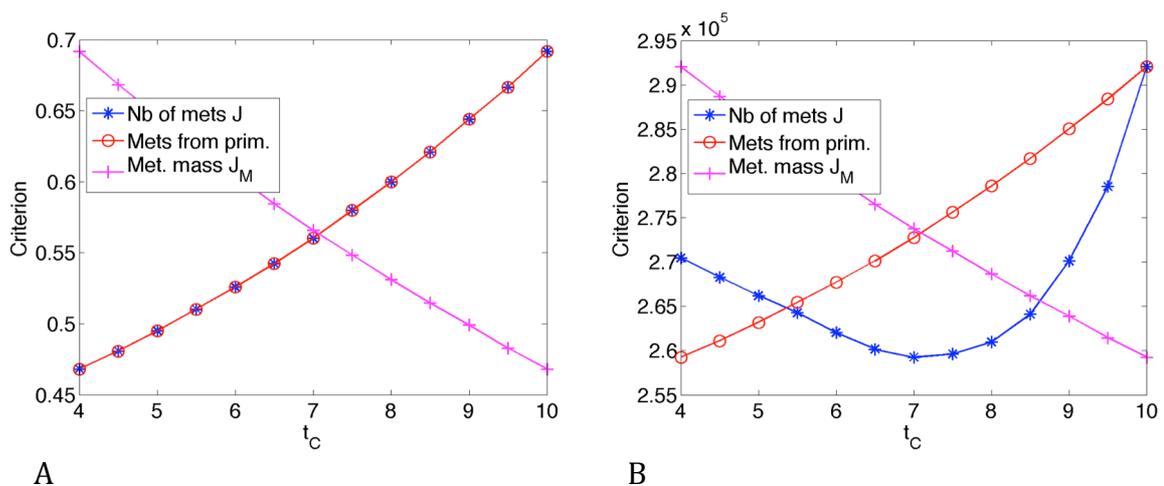

**Figure 4: Dependency on parameter m. For m=0.001, the curves for J and the metastases emitted by the primary tumor are identica m=0.001 B. m=100**

The change in the optimal strategy for the objective $J$ when varying $m$ seems to be specific to the CT monotherapy case and to the parameter $m$. We performed the same simulations in the AA monotherapy case and did not observe significant qualitative change in $J$, nor in $J_M$. Similarly, we did not obtain qualitative changes when varying the parameter $\alpha$, neither in the AA nor CT monotherapy case.

*CT-AA combination*

We turn now our interest on combination of an AA and a CT drug. The optimization problem is two-dimensional and the four objectives are now represented by surfaces shown in Figure 5.

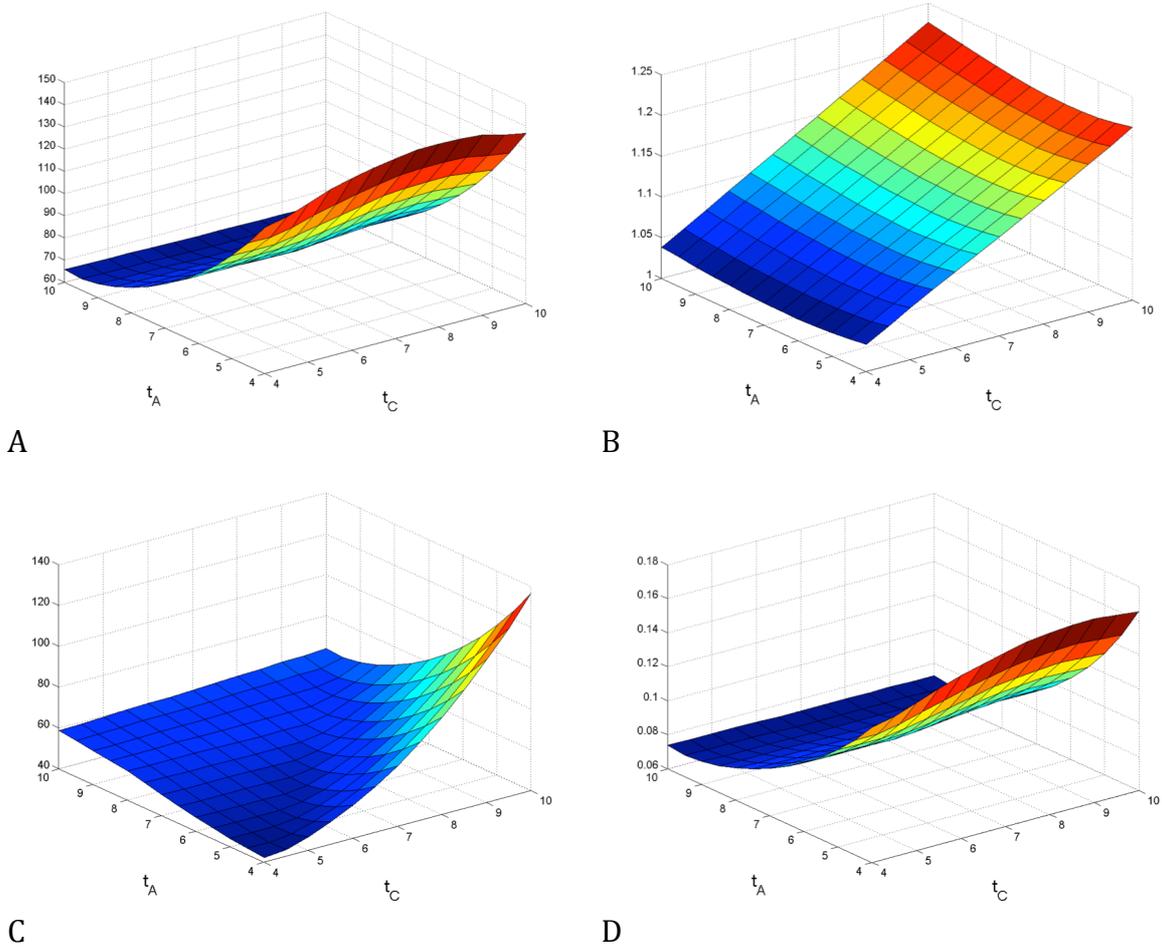

**Figure 5: Combination of a CT and an AA drug**

The optimal minimizer $(t_C^*, t_A^*)$ as well as the optimal values for the various objectives are given in the Table 2 and variation rates (see formula (3)) can be found in the Table 3.

| Objective | $J_m$ | $J_T$ | $J_T$ | $J_M$ |
|---|---|---|---|---|
| $(t_C^*, t_A^*)$ | (9.50,9.50) | (4.00,4.00) | (4.00,6.50) | (9.50,9.50) |
| **Optimal value** | 63.09 | 42.27 | 1.03 | 0.07 |

Table 2: Combination therapy. Minimizer ($t^*_C$, $t^*_A$) and optimal values

| Objective | $J_m$ | $J_T$ | $J_T$ | $J_M$ |
|---|---|---|---|---|
| **Variation rates** | -0.96/-0.87 | -0.94/-0.86 | 0.34/0.60 | -0.93/-0.83 |

Table 3: Variation rates for combination CT/AA

We observe again that depending on the objective, different strategies appear as optimal: metronomic for both drugs for the metastatic mass $J_M$ and the size at the end $J_T$, MTD for both drugs for the minimal size $J_m$, MTD for the CT and

$t_A^* = 6.5$ days for the number of metastases $J$. For all the objectives, the optimal strategy for both the CT and the AA is the same as in the monotherapy cases. This could suggest that there is no interplay between the actions of the two drugs, since adding another drug had no effect on the overall global strategy.

However, looking more precisely at what happens shows variability in the optimal solutions, as can already be seen on the $J_m$ surface of Figure 5. While $J$ has the shape of a surface representing the product of two functions $J(t_C, t_A) = f(t_C)g(t_A)$, it is not the case for $J_m$. To investigate this fact closer, we studied the following situation: we fixed the way of administering one drug, for

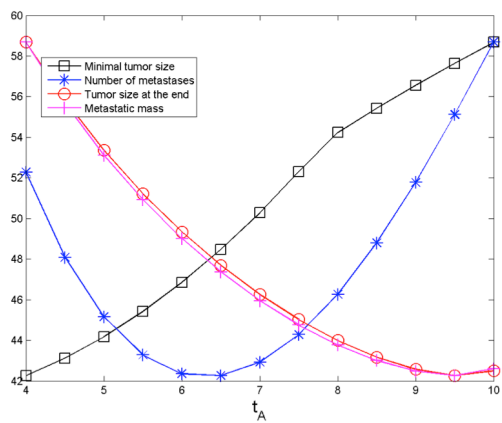

A. $t_C = 4$

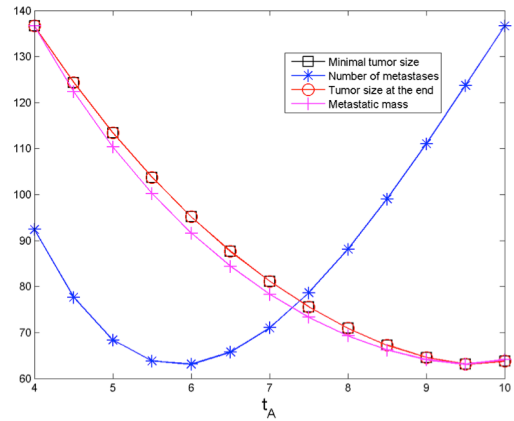

B. $t_C = 10$

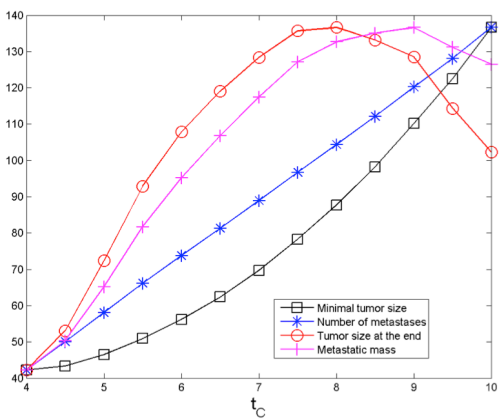

C. $t_A = 4$

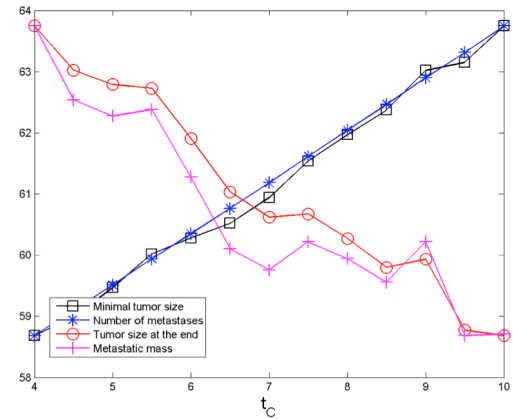

D. $t_A = 10$

Figure 6: Projections of the surface in the Figure 5 on the extremal planes

example imposed by the clinician, and looked at the effect on the optimal strategy for the other drug. This corresponds to look at projections of the surfaces from Figure 5 on the planes $t_C =$ constant or $t_A =$ constant. The extremal cases are shown in Figure 6.

In Figures 6.A and B, we fix the way of giving the CT drug and observe then what is the best strategy for the AA agent. The qualitative shape of objectives $J$, $J_T$ and

$J_M$ are almost identical in the two opposite cases for $t_C$ and the same as Figure 2, although optimal value for $J$ slightly moved to the right. Hence, adjunction of the chemotherapy did not change much the effects of the AA drug on these objectives, and varying the CT strategy does not impact. On the opposite, objective $J_m$ drastically changes, going from an increasing function to a decreasing one (notice that the curves for $J_T$ and $J_m$ are identical in Figure 6.B, indicating that the minimal size on $[0,T]$ is reached at the end time). With a MTD CT, it recommends MTD AA whereas with metronomic CT preference is given to metronomic AA.

Surprisingly, things are different when we fix the AA strategy and look at the impact on the CT one (Figures 6.C and D). While $J_T$ and $J_M$ suggested the metronomic strategy in the CT monotherapy case, they recommend now the opposite one. When the AA strategy is changed from MTD to metronomic, then $J_T$ and $J_M$ are again minimized by the metronomic strategy.

To illustrate further these concerns, we plot in Figure 7 the optimal AA strategy $t_A^*$ as a function of the CT strategy $t_C$ and conversely, that is, graphs of the maps $t_C \mapsto \underset{t_A}{\operatorname{argmin}} \Psi(t_A, t_C)$ on the left and $t_A \mapsto \underset{t_C}{\operatorname{argmin}} \Psi(t_A, t_C)$ on the right, for $\Psi = J_T, J_m, J, J_M$.

We observe an absence of symmetry between the two plots. This indicates that

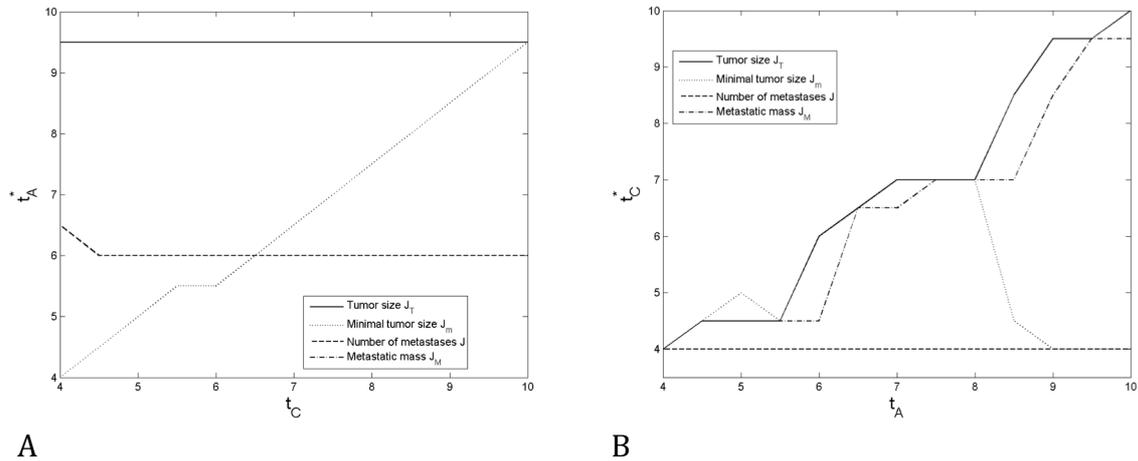

A                                      B

**Figure 7: Dependency of the optimal strategy for one drug on the strategy for the other one. A. Optimal AA strategy versus CT strategy. B. Optimal CT strategy versus AA strategy**

in the model AA and CT therapies have different dynamical effects, since varying the scheduling strategy does not impact the same. Focusing on the left curves, while $J$, $J_T$ and $J_M$'s optimal AA strategies are not affected by varying the CT scheduling, the $J_m$ curve is almost the identity line. This interesting fact suggests that both drugs should be ``synchronised'' in the sense that they should be given in the same way, with the same scheduling strategy. On the right plot, we also observe this synchronisation effect, but now for objectives $J_M$ and $J_T$.

**Conclusion and discussion**

The problem of optimizing the scheduling of the drugs in an anti-cancer therapy is of fundamental importance in the clinic. While reduction of the primary tumor size is often the first main target of therapy, number and size of the metastases have to be taken into account, especially in an adjuvant setting (after surgery). Using our previously developed phenomenological model for development of the metastatic population, we have defined an optimal control problem for the metastases. We then numerically studied the problem in a simplified case which is two-dimensional, but still clinically relevant and allowing to compare scheduling strategies ranging from condensed MTD to spread out metronomics temporal administration protocols.

We compared two objectives on the primary tumor size, the size of the tumor at the end of the simulation $J_T$ and the minimal size reached during the simulation time interval $J_m$, and two objectives on the metastases, their total number $J$ and the metastatic mass $J_M$, in the AA and CT monotherapy cases, as well as in combination. We obtained differences of the qualitative behavior of the objectives, with three possibilities : increasing function, suggesting the MTD strategy, decreasing function, suggesting the metronomic strategy or non-monotonous convex function with nontrivial value of the optimal strategy. In the monotherapy cases, the objective $J$ was never found to correlate with $J_T$ nor $J_m$, thus emphasizing the relevance of adding a metastatic component in the optimal control problem of the drugs scheduling, since it says that the optimal strategy for reducing metastatic spreading is different from the primary tumor ones. Since all the objectives have different (and sometimes even opposite) behaviors, the natural question that arises is: which one has to be chosen? Maybe some suitable weighting of the objectives could be used. Another way of integrating both tumor size and number of metastases is to consider the metastatic mass $J_M$. For most of the cases, this objective has the same minimizer than $J_T$ which was heuristically explained because the relative reduction in the tumor size at the end between the worst and best scheduling strategy was higher than reduction of the number of metastases. For both the AA and the CT drug, it suggests delivering the drug at low rate for a large time. This finding corroborates with the metronomic approach for CT drugs that proved to give interesting results, in preclinical (37) and clinical (3–5) settings. This result was robust when varying the parameter $m$ of intrinsic metastatic aggressiveness of the patient's disease and suggests that regarding to the total metastatic burden, the optimal scheduling strategy leads to a situation with a lot but small metastases, which is preferred to a situation with fewer but bigger metastases.

In the combined therapy case, the qualitative behaviors of the objectives were all different again. Looking at the situation where the scheduling strategy is fixed for one drug and decision has to be made for the other drug, we observed an interesting synchronization effect suggesting that the best strategy for the second agent is the same as the fixed strategy used for the first one. This happened only for objective $J_m$ when fixing the CT strategy and for objectives $J$ and $J_T$ when fixing the AA strategy.

Overall, our results underline the importance of scheduling of anticancer agents and the necessity to take into account for the metastatic process in the design of treatment protocols. The objective to be optimized by the therapy should be precisely defined. In particular, there are differences in the optimal scheduling strategy between primary tumor objectives and metastatic ones. However, our results suggest superiority of the metronomic schedule on the classical MTD since it was observed in simulations to perform better for reduction of end tumor size and metastatic burden. These could be considered as desired objectives when thinking of long-term control of the cancer disease and not total eradication of all tumors.

Although the two-dimensional situation numerically studied here is already rich and complex, the numerical resolution of the complete infinite-dimensional optimal control problem on the number of metastases should also be addressed. The optimal control problem on the metastases is not linear but rather bilinear in $(u,\rho)$. Resolution of such an optimization problem is not standard. Without resolving the complete optimal control problem, we could also investigate slightly more elaborated situations yet still simple and suboptimal, for example by dividing the time interval in two or more and applying what we did on the whole interval to each sub interval.

On the modeling part, the optimal control problem that we defined is not completely clinically relevant since the metastatic problem typically arises on larger time scales, for example in determining the best way to avoid relapse after surgery. Since it is not numerically neither clinically tractable to compute/administrate a continuous control on a very large time interval, we could impose some periodic structure that remains to be specified. If we still focus on optimizing metastatic emission and growth on the time scale of a therapy cycle (for example, 21 days), then we should integrate more complex modeling of hematotoxicities of the chemotherapy, as done in the MODEL I project (23,24). For larger time frames, drug resistance should also been taken into account. Substantial literature exists about mathematical modeling of resistance to chemotherapy due to genetic mutations, ranging from stochastic models dealing with maximization of the probabilities of no development any resistant phenotype and cure (see for instance the seminal work of Goldie and Coldman (49)) to differential equations-based models using compartments (40,22,48,52). Cellular efflux mechanisms inducing drug resistance could also be considered, as done in (56). Most of existing models are focused on cytotoxic agents, known to induce resistance in the genetically unstable cancerous cells population. Anti-angiogenic therapy was thought to be exempted of resistances because it is directed against the more genetically stable endothelial cells and was even called a therapy resistant to resistance (50). However, more recent findings suggest otherwise (51) and call for a different philosophy in the modeling of AA therapy drug resistance.

Interactions between AA and CT therapy could also be more precisely dealt with, for instance by considering a weighted sum of tumor volume and carrying capacity in the tumor objective as done in (48). Moreover non-trivial biological interactions between the two types of drug could be taken into account. Indeed, drug delivery is dependent on vascular supply that is negatively affected by the AA therapy. On the other hand it has been proposed a normalization effect (44) of AA drugs that would result in vasculature pruning, improving the poor quality of neo-angiogenic tumor blood vessels and thus delivery of the molecular agents. Mathematical models that address these concerns have been proposed (45,46) and could be used to refine the modeling.

## Acknowledgements

We thank Assia Benabdallah for helpful discussions on the theoretical part. This work was supported by the National Cancer Institute under Award Number U54CA149233 (to L. Hlatky). The content is solely the responsibility of the authors and does not necessarily represent the official views of the National Cancer Institute or the National Institutes of Health.